\documentclass[aps,prb,superscriptaddress,twocolumn,floatfix]{revtex4-2}
\usepackage{units}
\usepackage{amsmath}
\usepackage{amssymb}
\usepackage{graphicx}
\usepackage{bm}
\usepackage{multirow,color,relsize,ulem,microtype,tabularx}

\newcommand{\be}{\begin{equation}}
\newcommand{\ee}{\end{equation}}

\newcommand{\pt}{\cal PT}

\begin{document}

\title{Analysis of Dirac exceptional points and their isospectral Hermitian counterparts}

\author{Jose H. D. Rivero}
\affiliation{Department of Materials Science and Engineering, University of Pennsylvania, Philadelphia, PA 19104, USA}

\author{Liang Feng}
\affiliation{Department of Materials Science and Engineering, University of Pennsylvania, Philadelphia, PA 19104, USA}

\author{Li Ge}
\email{li.ge@csi.cuny.edu}
\affiliation{\textls[-18]{Department of Physics and Astronomy, College of Staten Island, CUNY, Staten Island, NY 10314, USA}}
\affiliation{The Graduate Center, CUNY, New York, NY 10016, USA}

\begin{abstract}
Recently, a Dirac exceptional point (EP) was reported in a non-Hermitian system. Unlike a Dirac point in Hermitian systems, this Dirac EP has coalesced eigenstates in addition to the degenerate energy. Also different from a typical EP, the two energy levels connected at this Dirac EP remain real in its vicinity and display a linear instead of square root dispersion, forming a tilted Dirac cone in the hybrid space consisting of a momentum dimension and a synthetic dimension for the strength of non-Hermiticity. In this report, we first present simple three-band and two-band matrix models with a Dirac EP, where the linear dispersion of the tilted Dirac cone can be expressed analytically. Importantly, our analysis also reveals that there exist Hermitian and non-Hermitian systems that have the same (real-valued) energy spectrum in their entire parameter space, with the exception that one or more degeneracies in the former are replaced by Dirac EPs in the later. Finally, we show the existence of an imaginary Dirac cone with an EP at its center.

\end{abstract}

\maketitle

\section{Introduction}

Dirac points, the degeneracies at the center of Dirac cones, give graphene \cite{Novoselov,Geim} and other Dirac matters \cite{Wehling} their unusual electronic properties. Their signature, i.e., the linear dispersion around a Dirac point in momentum space, signals a massless fermion that differs significantly from a classical object with the quadratic kinetic energy relation. Similar to all other degeneracies in Hermitian systems, a Dirac point corresponds to two different quantum eigenstates at the same energy level and hence is an example of a diabolic point \cite{Berry}.       

An exceptional point (EP) \cite{Heiss,EP}, on the other hand, features coalesced eigenstates at a degenerate energy level. Its existence is a unique feature of non-Hermitian systems, which arise when a physical system is represented by a Hamiltonian with partial degrees of freedom or by another operator that describes the exchange of energy or particles with the environment (such as the scattering matrix \cite{Chong1,Lin,Ge_PRA2012,Ge_PRA2015,Sweeney}). This approach has been adopted in studies ranging from nuclear decay \cite{Gamow} to photon lifetime in optical microcavities \cite{NPReview}, yielding insightful results as well as providing new directions of research \cite{NPReview,NPhyReview,RMP,Bender}. For example, the motion of eigenfrequencies or resonances in the complex plane can be quite unusual in the vicinity of an EP, which results in intriguing behaviors such as gain-suppressed lasing \cite{GeGain2011,Liertzer,ElGanainyLoss2014,BrandstetterLoss2014} and loss-induced lasing \cite{Peng}.

While a degeneracy in a Hermitian system cannot be an exceptional point, a degeneracy in a non-Hermitian system is not necessarily an EP. Such a non-EP degeneracy can occur accidentally or by symmetry, similar to the mechanisms in a Hermitian system. Some examples include the zero modes in a non-Hermitian flatband \cite{Ge_PRA2015b} and Dirac points constructed in non-Hermitian lattices \cite{Ge_PRJ2018,Chong}. %In these examples, the energy eigenvalues display a linear dependency on a system parameter away from the degeneracy, either in their real or imaginary parts. 
For the degeneracies that are indeed EPs, traditionally one associates their perturbative dependence on a system parameter with a fractional exponent [e.g., a square root for an EP of order 2 (EP2)], which implies a stronger response and hence potentially enhanced sensitivity \cite{EPsensing0,EPsensing1,EPsensing3}. Furthermore, if an EP has its roots in a non-Hermitian symmetry, such as parity-time ($\pt$) \cite{Bender,Makris,Guo,Ruter}, anti-$\pt$ \cite{antiPT0,antiPT1,antiPT2,antiPT3,antiPT4}, or particle-hole \cite{Malzard,zeromodeLaser,NHFlatband_PRL,Kawabata2019} symmetries, the non-Hermitian system experiences a corresponding spontaneous symmetry breaking or restoration across the EP. This phenomenon is usually accompanied by dramatic changes in the energy spectrum: It may transition from real to complex or even imaginary and vice-versa, which is used as one experimental signature of the EP itself \cite{antiPT1,Bittner,Ding,Schindler}.

Therefore, the recent, accidental finding of a Dirac EP \cite{DiracEP0} in a one-dimensional periodic $\pt$-symmetric system came as a surprise: Two energy bands connected by this Dirac EP display a linear and conical ``dispersion'' in a two-dimensional hybrid space, as a function of both momentum and the non-Hermitian parameter given by the optical gain and loss strength. In addition, the two energy bands remain real in the vicinity of the Dirac EP, without undergoing a spontaneous symmetry breaking. While a three-band matrix model was introduced in Ref.~\cite{DiracEP0} to capture the latter, the linear and conical dispersion at this Dirac EP remains to be elucidated. The differences between Dirac points, Dirac EPs, and conventional EPs are summarized in Table.~\ref{table1}. In the last row, state flip is marked ``possible'' instead of ``yes'' when a conventional EP2 is encircled, due to the non-adiabatic transition from the low-gain/high-loss state to the high-gain/low-loss state \cite{EP}. Such non-adiabatic behaviors caused by the different modal gain (or loss) are absent in the vicinity of a Dirac EP thanks to its real spectrum.

\begin{table}[htb]
\renewcommand{\arraystretch}{1.2}%
\footnotesize
\caption{Comparison of three types of degeneracies.}
\begin{center}
\begin{tabular}{m{4cm} m{1cm} m{1cm} m{2cm}}
%\begin{tabularx}{\linewidth}{  >{\centering\arraybackslash}X  >{\centering\arraybackslash}X  >{\centering\arraybackslash}X  >{\centering\arraybackslash}X}
\hline\hline
&    Dirac Points    & Dirac EPs   & Conventional EP2s \\ \hline
Algebraic multiplicity &  2  &  2  &  2 \\ \hline
Geometric multiplicity &  2  &  1  &  1 \\ \hline
Coalescence of wave functions &  No & Yes & Yes \\ \hline
Locally real spectrum & Yes & Yes & No\\ \hline
Branch cuts at degeneracy &  No &  No &  Yes \\ \hline
Node type in energy (real part) &  Point &  Point &  Line or surface \\ \hline
Energy dispersion &  Conical & Conical & Square roots \\ \hline
State flip when encircled &  No & No &  Possible \\  \hline\hline
%\end{tabularx} 
\end{tabular}
\label{table1}
\end{center}
\end{table}

In this Report, we first introduce a revised three-band model and show explicitly the linear dispersion of the (tilted) Dirac cone centered at the Dirac EP, via a perturbative treatment. Our approach is different from the standard procedure of using the alternating Puiseux series for a conventional EP \cite{Pick,Puiseux}, which is inapplicable at a Dirac EP as we will show. Because the Dirac EP connects two bands instead of three, we further reduce this three-band model to a two-band non-Hermitian Hamiltonian where the Dirac EP can be found. Using a slightly different approach, we find, interestingly, that while the same linear and conical dispersion is produced in the hybrid dimensions, the Dirac point becomes a diabolic point instead, with two distinct eigenstates. We show that this dilemma can be resolved by revisiting the three-band model. More importantly, this comparison has a far-reaching implication: There exist Hermitian and non-Hermitian systems that have the same (real-valued) energy spectrum in their entire parameter space, with the exception that one or more degeneracies in the former are replaced by Dirac EPs in the later. Finally, we show the existence of an imaginary Dirac cone with an EP at its center.

\section{Three-band model}

The systems exhibiting a Dirac EP we study originate from the following Schr\"odinger equation
\be
i\frac{d}{dt}\psi(x,t) = [-\partial_x^2 + V(x)]\psi(x,t), \label{eq:Schrodinger}
\ee
where time, position, and potential are in their dimensionless forms and $\hbar=1$. $V(x)=V_0(\cos\,x+i\tau\sin\,x)\,(\tau\geq0)$ is a complex potential with the spatial period $a=2\pi$. It is $\pt$-symmetric and satisfies $V(x)=V^*(-x)$ \cite{NPReview}. The asterisk denotes complex conjugation and represents time reversal, and the imaginary part of the potential represents optical gain and loss with strength $\tau$ \cite{NPReview}. 

The Dirac EP in this system is found at the point contact between the second and third bands in the first Brillouin zone, where $\tau=1$ and $k=0$. To gain analytical understanding of this Dirac EP, Ref.~\cite{DiracEP0} first expanded the Bloch wave function in the plane-wave basis, i.e.,
\be
\psi_{nk}(x,t) =  e^{ikx-i\omega t}\sum_{m\in\mathbb{Z}} a_m e^{imx},
\ee
and derived the Bloch Hamiltonian 
\begin{align}
H_k = \sum_{m\in\mathbb{Z}} (m+k)^2|m\rangle\langle m| \;&+\; t_- |m\rangle\langle m+1|\nonumber\\
 &+\; t_+ |m\rangle\langle m-1|\label{eq:Hk}
\end{align}
that satisfies 
\be
H_k\Psi_{nk}(m) = \omega_{nk} \Psi_{nk}(m). 
\ee
Here $n=1,2,\ldots$ is the band index, $\omega_{nk}$ is energy of the $n$th band when the momentum is $k$, $\Psi_{nk}(m)=[\ldots,a_{-1},a_0,a_1,\ldots]^{T}$is the wave function in the plane-wave basis, and $t_\pm=V_0(1\pm\tau)/2\in\mathbb{R}$. Then by truncating $H_k$ given by Eq.~(\ref{eq:Hk}) and keeping only the $m=-1$, $0$ and 1 block, Ref.~\cite{DiracEP0} showed analytically that indeed the second and third bands remain real-valued when we increase $\tau$ across $1$, i.e., the system does not experience a spontaneous $\pt$ breaking by going through the Dirac EP. While this analysis was only performed for $k=0$ and the truncation turns out to be a crude approximation, it shone light on how a Dirac EP could be constructed. 

Below we first use the insight from this truncation to introduce a revised three-band model, where a Dirac EP exists and where the linear dispersion of its (tilted) Dirac cone can be expressed analytically. In this model, we allow $k$ to be a (small) free parameter in addition to the gain and loss strength $\tau$, and the three-band non-Hermitian Hamiltonian is given by
\be
H^{(3)} = 
\begin{pmatrix}
1-2k & t_- & 0 \\
t_+ & 0 & t_- \\
0 & t_+ & 1+2k
\end{pmatrix}\label{eq:H3}
\ee
with the asymmetric couplings $t_\pm$ introduced in Eq.~(\ref{eq:Hk}). The eigenvalues $\omega_i$ of $H^{(3)}$ are the solutions of the characteristic polynomial 
\be
\omega_i(1-\omega_i)^2 + 2t^2(1-\omega_i) - 4k^2\omega_i = 0, \label{eq:charateristicPoly}
\ee
where $t^2\equiv t_-t_+=V_0^2(1-\tau^2)/4$. This cubic equation can be solved analytically, but the resulting expressions for $\omega_i$'s are rather complicated (e.g., with a square root inside a cubic root) and do not help us understand the properties of the Dirac EP. We could perform a Taylor expansion of these expressions for $\omega_i$'s, but a much simpler approach is to expand Eq.~(\ref{eq:charateristicPoly}) directly, which gives the same results. We do note that being a cubic equation with real coefficients, Eq.~(\ref{eq:charateristicPoly}) indicates that the three energy bands are either all real or one real plus a complex conjugate pair. Therefore, it does not exclude the possibility of a spontaneous $\pt$ breaking, which nevertheless does not take place at the Dirac EP. 

\begin{figure}[tb]
\includegraphics[clip,width=\linewidth]{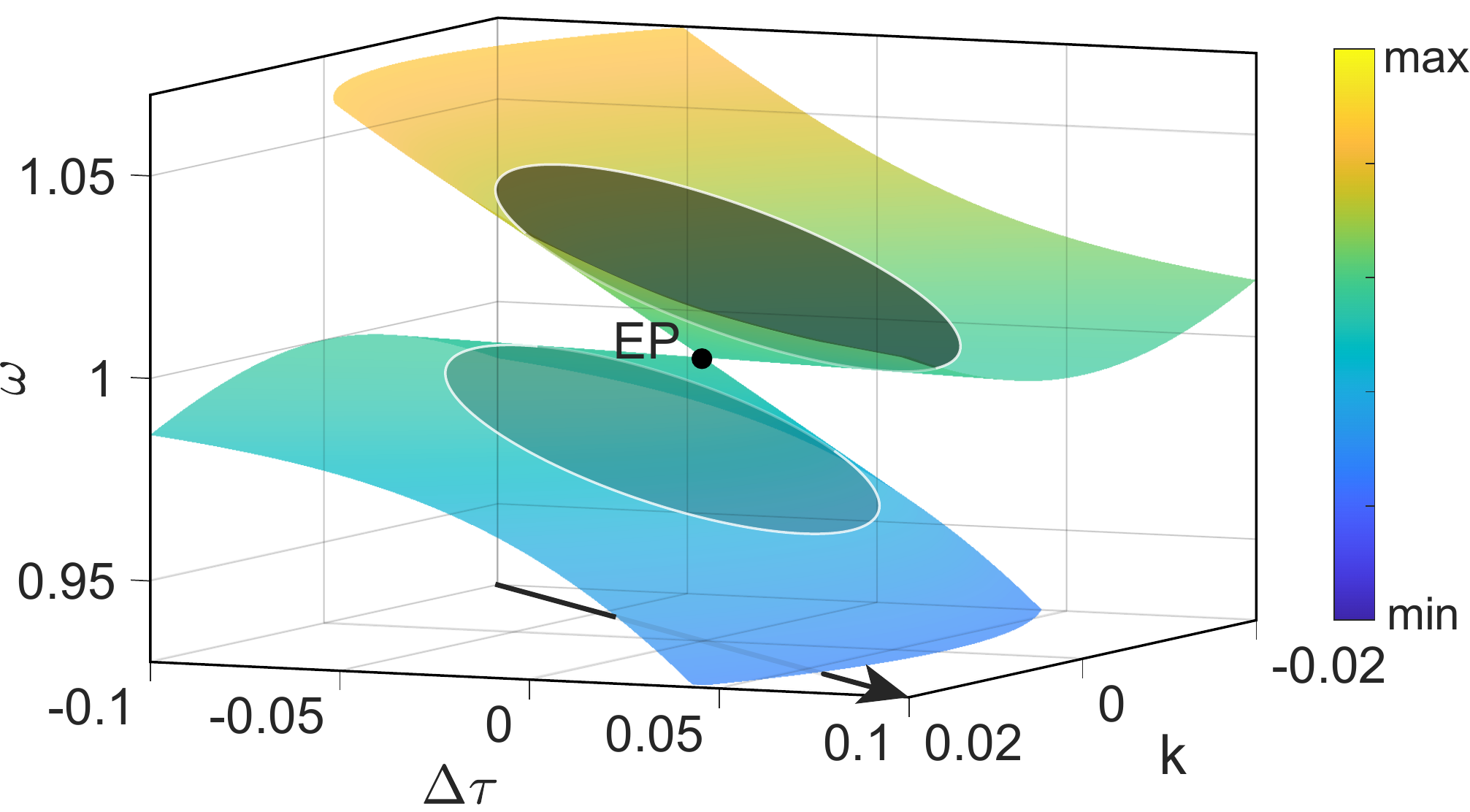}
\caption{Tilted Dirac cone of $H^{(3)}$ centered at a Dirac EP. $k$ is momentum and $\Delta\tau$ is the change of the non-Hermitian strength from its value (i.e., 1) at the Dirac EP. $V_0=1$ is used in the couplings $t_\pm$. The intersection of the Dirac cone and two parallel planes given by $\omega=1-s\Delta\tau\pm d$ are also shown, where $\omega_0=1$ is the energy at the Dirac EP, $s=\sqrt{2}-1$, and $d=1/40$. The arrow shows the diagonal direction used in Fig.~\ref{fig:dispersion}(b).  
} \label{fig:DiracEP}
\end{figure}

It is straightforward to verify that this EP has energy $\omega=1\equiv\omega_0$ (again $\hbar=1$) and exists at $\tau=1,k=0$. Its coalesced eigenstate is given by $[a_{-1},a_0,a_1]^{T}=[0,0,1]^T\equiv\Psi_0$. This EP is the point contact of the second and third bands that form a tilted Dirac cone (see Fig.~\ref{fig:DiracEP}), similar to the original Hamiltonian $H_k$ shown in Ref.~\cite{DiracEP0}. To derive the dispersion relation near this Dirac EP, we write $\omega_i \equiv \omega_0+\Delta\omega_i\,(|\Delta\omega_i|\ll\omega_0)$ and study how $\Delta\omega_i$ depends on the two small parameters $\Delta\tau\equiv\tau-1$ and $k$. 

Here we take the advantage of knowing that the dispersion is linear and conical, which implies that $\Delta\omega_i$ is of the same order as $k$ and $\Delta\tau$. Consequently, Eq.~(\ref{eq:charateristicPoly}) becomes
\be
\Delta\omega_i^2 - 2t^2\Delta\omega_i - 4k^2 = 0,\label{eq:expansion}
\ee 
to the leading order of $\Delta\omega_i$ (i.e., $\Delta\omega_i^2$), where $t^2\approx -(V_0^2/2)\Delta\tau$ is also of the same order as $\Delta\tau$ (with $V_0$ chosen to be of order unity). 
We then find
\be
\Delta\omega_i = t^2 \pm \sqrt{t^4+4k^2} \label{eq:DeltaOmega}
\ee
or
\be
\Delta\omega_i \approx (-\alpha\pm\sqrt{4+\alpha^2})k \label{eq:lineardispersion}
\ee  
in any arbitrary direction $\Delta\tau = (2\alpha/V_0^2)k\,(\alpha\in\mathbb{R})$ from the origin of the $\Delta\tau$-$k$ plane. 

\begin{figure}[b]
\includegraphics[clip,width=\linewidth]{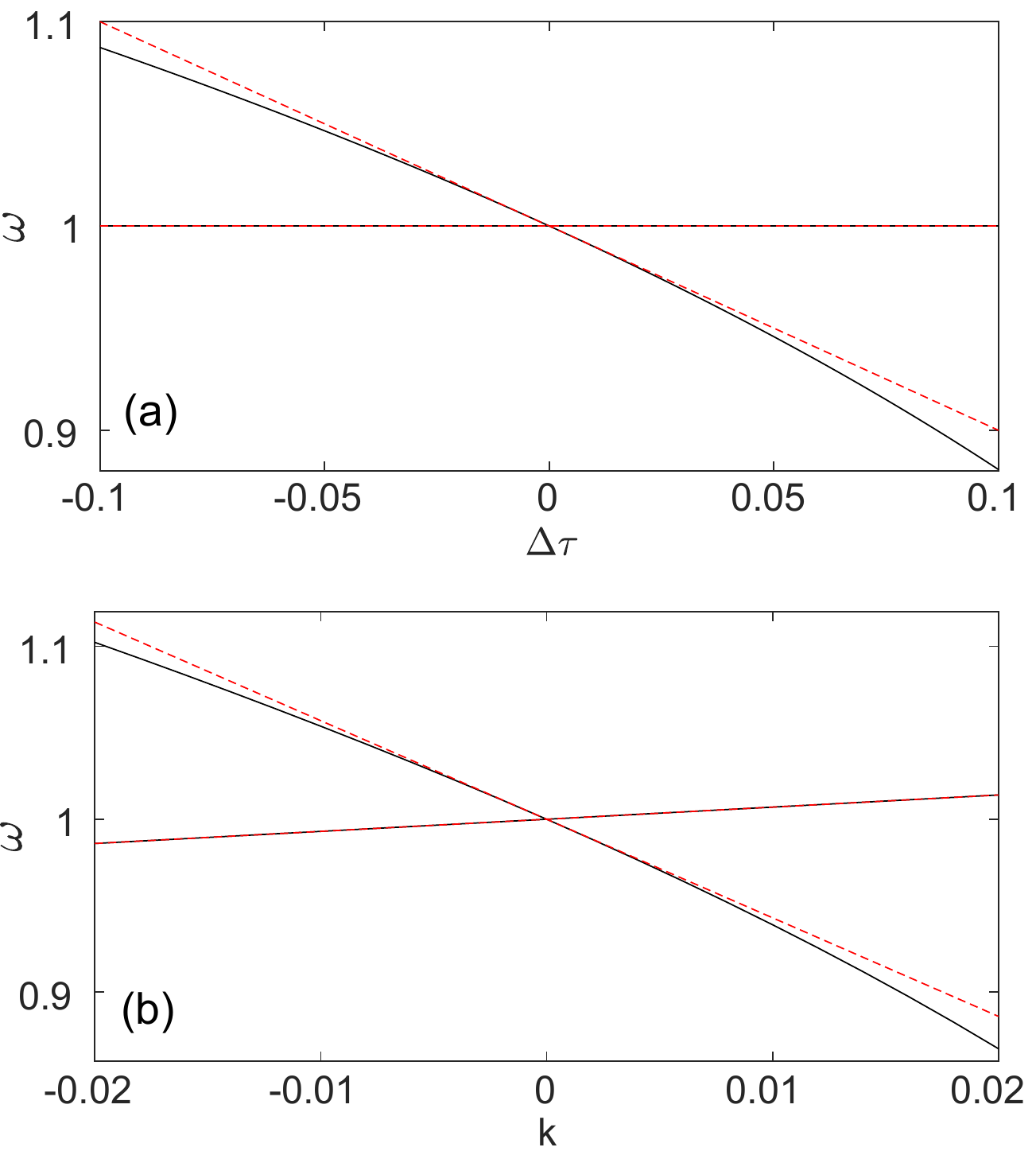}
\caption{Linear dispersion of the tilted Dirac cone along (a) $\Delta\tau$ and (b) the diagonal direction given by $\Delta\tau=5k$ in Fig.~\ref{fig:DiracEP}. Solid and dashed lines show the actual and approximated band energies, which cannot be distinguished by eye for the band closer to $\omega=1$.} 
\label{fig:dispersion}
\end{figure}

The square root in Eq.~(\ref{eq:lineardispersion}) directly captures the linear and conical dispersion of the Dirac cone shown in Fig.\ref{fig:DiracEP}, with the linear term in front of it explaining the tilt of this Dirac cone. We also note that this expression is exact when $\Delta\tau=0$ (i.e., along the $k$ direction from the Dirac EP), which can be checked easily by setting the second term in Eq.~(\ref{eq:charateristicPoly}) to be zero. In Fig.~\ref{fig:dispersion}, we show the comparison of Eq.~(\ref{eq:lineardispersion}) and the actual band energies along two other directions, i.e., the $\Delta\tau$ axis and the one with $\alpha=2.5$ (diagonal direction in Fig.~\ref{fig:DiracEP}). We observe that the band closer to $\omega=1$ is better approximated by Eq.~(\ref{eq:lineardispersion}), which will be explained using the equivalent two-band model below where the analysis becomes easier. 

We also note that our approach is different from the perturbative expansion used at a conventional EP, which utilizes alternating Puiseux series \cite{Puiseux} with fractional powers of a small parameter. In fact, this standard treatment is inapplicable at a Dirac EP as we show below. This approach calculates the perturbative corrections to the eigenstates and their energies when the Hamiltonian changes from $H_0$ to $H = H_0 + \epsilon H_1\,(\epsilon\ll1)$, and when an EP is the result of two coalescing eigenstates (i.e., of multiplicity 2) \cite{Pick}, the alternating Puiseux series assume integer and half-integer powers of $\epsilon$, i.e.,
\begin{align}
\omega_\pm & = \omega_0 \pm \epsilon^{1/2}\omega_1 + \epsilon \omega_2 \pm \epsilon^{3/2}\omega_3 + \ldots \label{eq:Puiseux}\\
\Psi_\pm & = \Psi_0 \pm \epsilon^{1/2}\Psi_1 + \epsilon \Psi_2 \pm \epsilon^{3/2}\Psi_3 + \ldots
\end{align}
Here $\omega_\pm$ are the energies of the two eigenstates $\Psi_\pm$ that become coalesced at the EP, with energy $\omega_0$ and wave function $\Psi_0$. Even if we ignore the $\epsilon^{1/2}$ term, Eq.~(\ref{eq:Puiseux}) indicates clearly that $\omega_\pm$ have the same linear dependence on $\epsilon$. In other words, they stay the same to the linear order and hence cannot form a Dirac cone in this perturbative expansion. %Therefore, the perturbative expansion with alternating Puiseux series is inapplicable at a Dirac EP. 

\section{Two-band model}

Since the Dirac EP only connects two bands of $H^{(3)}$ [i.e., the second and third bands; represented by the two signs in Eq.~(\ref{eq:DeltaOmega})], it should be possible to reduce $H^{(3)}$ to a two-band Hamiltonian, which not only simplifies the understanding of the Dirac EP but also provides another instance where it exists. To this end, we first write down the time-dependent Schr\"odinger equation corresponding to $H^{(3)}$:
\begin{align}
i\frac{d}{dt} a_{-1} &= (1-2k) a_{-1} + t_-a_0 \label{eq:1}\\
i\frac{d}{dt} a_{0} &= t_+ a_{-1} + t_-a_1 \label{eq:2}\\
i\frac{d}{dt} a_{1} &= (1+2k) a_1 + t_+a_0 \label{eq:3}
\end{align}
Using $a_0\propto e^{-i\omega_i t}$ in an eigenstate with energy $\omega_i$, we obtain 
\be
a_0 = (t_+ a_{-1} + t_-a_1)/\omega_i
\ee
from Eq.~(\ref{eq:2}) and use this expression to eliminate $a_0$ in Eqs.~(\ref{eq:1}) and (\ref{eq:3}). The result is a two-band Hamiltonian
\be
H_a = (1+t^2/\omega_i)\bm{1}+\begin{pmatrix}
-2k & t_-^2/\omega_i \\
t_+^2/\omega_i & 2k
\end{pmatrix}, \label{eq:Ha}
\ee
where $\bm{1}$ is the identity matrix. Because no approximations have been used in deriving $H_a$, it has the same eigenvalues $\omega_i$'s as $H^{(3)}$. Furthermore, we note that the eigenvalue $\omega_i$ appears in the Hamiltonian $H_a$ itself, and hence this problem can be treated as a nonlinear eigenvalue problem \cite{NLE}. Nevertheless, if we replace $\omega_i$ by $\omega_0=1$ in $H_a$ itself near the expected EP, i.e.,
\be
H_a \rightarrow H_a' = (1+t^2)\bm{1}+
\begin{pmatrix}
-2k & t_-^2 \\
t_+^2 & 2k
\end{pmatrix}, \label{eq:Ha1}
\ee
we immediately find that $H_a'$ takes the Jordan normal form when $k=0$ and $\Delta\tau=0$, at which $t_-$ vanishes as well that leads to an EP with energy $\omega_0=1$. It is easy to check that the two eigenvalues of $H_a'$ are the same as those given by Eq.~(\ref{eq:DeltaOmega}), and hence it gives the same linear and conical dispersion relation (\ref{eq:lineardispersion}). 

If we have not replaced $\omega_i$ by $\omega_0=1$ in Eq.~(\ref{eq:Ha}), we can also express the nonlinear eigenvalues of $H_a$ in a self-consistent way, i.e.,
\be
\omega_i = (1+t^2/\omega_i) \pm \sqrt{4k^2+t^4/\omega_i^2},\label{eq:NLE}
\ee  
which is equivalent to Eq.~(\ref{eq:charateristicPoly}). It is then clear that the errors introduced by taking $\omega_i\approx\omega_0$ in Eq.~(\ref{eq:Ha1}) only originate from the $t^2$, $t^4$ terms in Eq.~(\ref{eq:NLE}). They are partially cancelled (increased) in the solution with the ``$-$'' (``$+$'') sign in Eq.~(\ref{eq:NLE}), which is also closer to (further from) the energy at the Dirac EP. This is exactly what we have observed in Fig.~\ref{fig:dispersion}. 

\section{Isospectral Hermitian and non-Hermitian systems}

As mentioned in the introduction, there is more than one way to reduce $H^{(3)}$ to a two-band Hamiltonian. Because the amplitudes $a_{\pm1}$ are coupled indirectly through $a_0$ in $H^{(3)}$, one may seek eigenstates in the forms of symmetric and antisymmetric superpositions of $a_{\pm1}$, weighted by the couplings $t_\mp$: 
\be
a_\pm = t_+a_{-1} \pm t_-a_1.
\ee 
Similar to the first approach above, we also eliminate $a_0$ and express it in terms of $a_\pm$. The resulting two-band Hamiltonian for $a_\pm$ is then found by multiplying Eqs.~(\ref{eq:1}) and (\ref{eq:3}) by $t_+,t_-$ respectively and taking the summation and difference of the results: 
\be
H_b = 
\begin{pmatrix}
1+2t^2/\omega_i & -2k \\
-2k & 1
\end{pmatrix}. \label{eq:Hb}
\ee
Again, by approximating $\omega_i$ in $H_b$ by $\omega_0=1$ at the Dirac EP, i.e.,
\be
H_b\rightarrow
H_b' = 
\begin{pmatrix}
1+2t^2 & -2k \\
-2k & 1
\end{pmatrix},
\ee
it is straightforward to show that the two eigenvalues of $H_b'$ are given by Eq.~(\ref{eq:DeltaOmega}), and we recover the linear and conical dispersion relation (\ref{eq:lineardispersion}). However, $H_b'$ (and $H_b$) becomes an identity matrix at where we expect to find the Dirac EP, i.e., $k=\Delta\tau=0$, and hence the degeneracy $\omega_0=1$ is a diabolic point instead of an EP, with two distinct eigenstates $[0,1]^T$ and $[1,0]^T$. This should not be surprising, however, because $H_b'$ is a real symmetric matrix, and hence it is Hermitian and cannot have an EP. 

This apparent contradiction can be easily resolved by realizing that one step leading to $H_b$ [i.e., multiplying Eq.~(\ref{eq:3}) by $t_-$] fails when $t_-=0$ (or equivalently, $\Delta\tau=1$) where the degeneracy exists. Therefore, $H_b$ is not equivalent to $H_a$ at $\Delta\tau=1$ and it is allowed to  differ from $H_a$, i.e., having a diabolic point instead of an EP. 

More importantly, this comparison reveals a far-reaching implication: There exist Hermitian and non-Hermitian systems that have the same (real) energy spectrum in their \textit{entire} parameter space, with the exception that one or more degeneracies in the former are replaced by Dirac EPs in the latter. 

This observation holds for the pair of linear Hamiltonians $H_a',H_b'$ in the entire $\tau$-$k$ parameter space. Here $H_b'$ should be treated as given, and hence the illegitimacy mentioned above from $H^{(3)}$ to $H_b$ at the Dirac EP is irrelevant. Although there is only one degeneracy (a diabolic point) replaced by one EP in our examples here, cases with more or even all degeneracies replaced by EPs can be trivially constructed. For example, one can generate a series of two-band Hamiltonians similar to $H_a'$ but with different energy shifts (i.e., replacing $\omega_0=1$ in its diagonal elements by an increasing series $\Delta_m$'s) and then stack them to form a block diagonal non-Hermitian Hamiltonian $H_A'$. By following the same process but using $H_b'$ instead, we end up with another block diagonal Hamiltonian $H_B'$ which is Hermitian. It is easy to see that they have the same energy eigenvalues throughout the parameter space $\tau$-$k$, with $\Delta_m$'s being diabolic points in $H_B'$ but EPs in $H_A'$.

We note that the isospectral property between a Hermitian and a non-Hermitian system we report here is stronger than that found in Ref.~\cite{DiracEP0}, where this equivalence was only established in the $\pt$-symmetric regime of a linear non-Hermitian system and away from its EP. The more general claim here is made possible partly by the elimination of the conventional EPs of the system studied in Ref.~\cite{DiracEP0}, which occur at the edge of the Brillouin zone when $\tau=1$. Near these conventional EPs, the band energies become complex along the $\Delta\tau$ direction and hence lose the equivalence to their Hermitian counterparts. If we restrict our discussion to the one-dimensional parameter space along $k$ with $\tau$ fixed at 1, one may attempt to claim that this system also have the more general isospectral property reported here: Its entire band structure is real valued in the first Brillouin zone and identical to that of a Hermitian system with $V(x)=0$; its EPs, both the conventional ones at the band edge and the unconventional ones at the center of the Brillouin zone (including the Dirac EP), are replaced by degeneracies in the Hermitian system. However, one quickly realizes that with $V(x)=0$, this ``crystal'' is just free space with a single dispersion relation $\omega=k^2$. Therefore, its degeneracies in the band analysis are artifacts of applying the periodic boundary condition to a ``unit cell'' of an arbitrary length, resulting in the folding of this single energy relation into the first Brillouin zone.

\section{linear and conical dispersion}

\begin{figure}[hbt]
\includegraphics[clip,width=\linewidth]{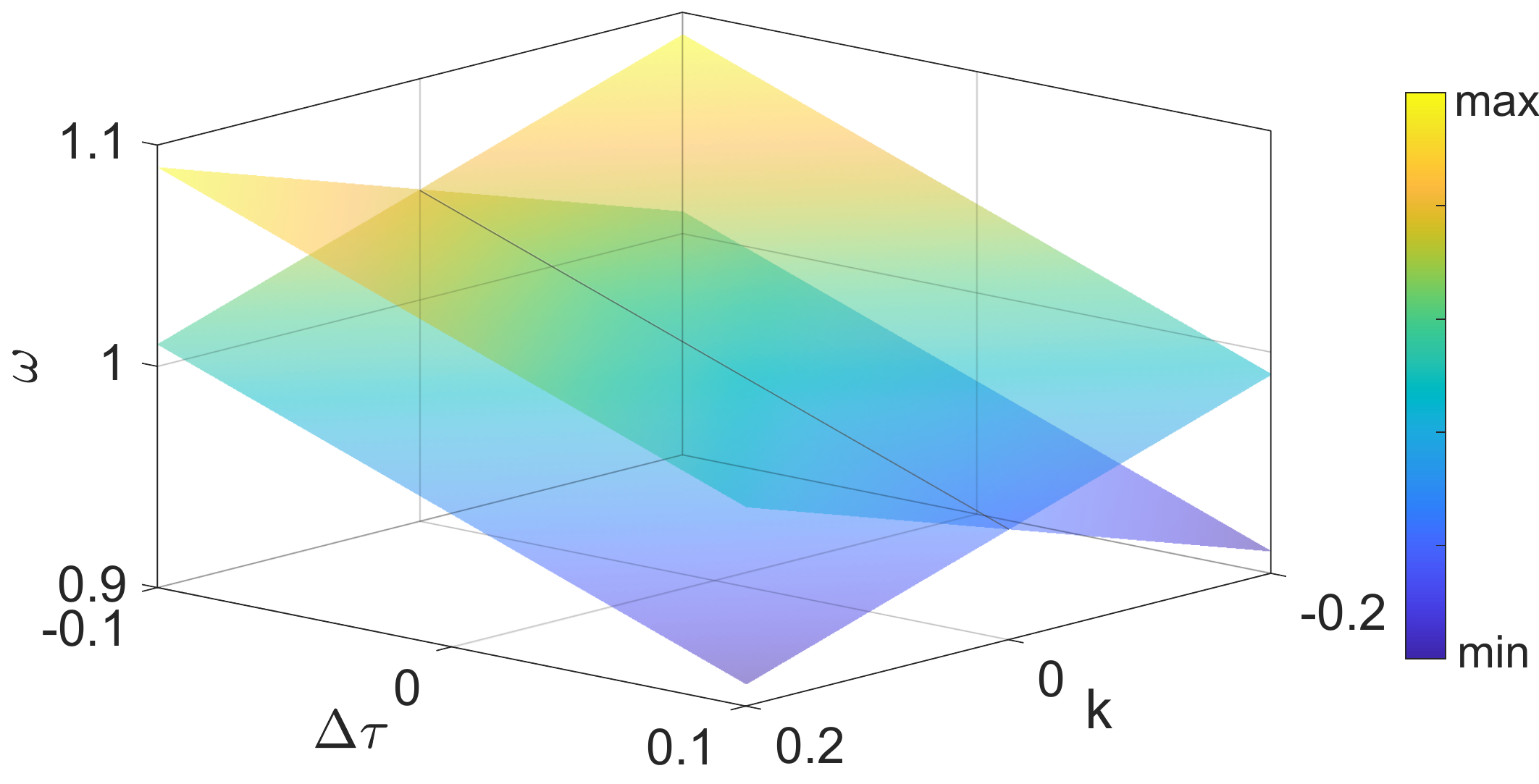}
\caption{Energy surfaces of $H_a''$ intersecting at a Dirac exceptional line instead of a Dirac EP. The parameters are the same as in Fig.~\ref{fig:DiracEP}. 
} \label{fig:TaylorEnergySurfaces}
\end{figure}

While both our three-band model $H^{(3)}$ and linearized two-band model $H_a'$ host a Dirac EP, there is a noticeable difference between them: The small changes in the former, i.e., $H^{(3)}=H_0+\Delta H$ where
\be
H_0 =  
\begin{pmatrix}
1 & 0 & 0\\
V_0 & 0 & 0 \\
0 & V_0 & 1
\end{pmatrix},\quad
\Delta H =  
\begin{pmatrix}
-2k & -g & 0\\
g & 0 & -g \\
0 & g & 2k
\end{pmatrix},\label{eq:H1_H3}
\ee
and $g\equiv V_0\Delta\tau/2$, are of the same order and linear in terms of $k$ and $\Delta\tau$; there are, however, higher-order terms proportional to $\Delta\tau^2$ in the two-band model $H_a'=H_0+\Delta H+\Delta H'$, where
\be
H_0 =  
\begin{pmatrix}
1 & 0 \\
V_0^2 & 1
\end{pmatrix},
\quad
\Delta H =  
-\frac{V_0^2}{2}\Delta\tau\bm{1} + 
\begin{pmatrix}
-2k & 0 \\
V_0^2\Delta\tau & 2k
\end{pmatrix},\nonumber
\ee
and
\be
\Delta H'=
\frac{V_0^2}{4}\Delta\tau^2
\begin{pmatrix}
-1 & 1\\
1 & -1
\end{pmatrix}.\nonumber
\ee
Without these higher-order terms, especially the upper right element in $\Delta H'$, $\omega_0=1$ is still an EP of the resulting Hamiltonian 
$H_a'' \equiv H_0 + \Delta H$, and the dispersion at this EP is still linear:
\be
\Delta\omega_i = -\frac{V_0^2}{2}\Delta\tau \pm 2k. \label{eq:2band}
\ee
However, these two energy surfaces intersect at an EP line instead of a Dirac EP (see Fig.~\ref{fig:TaylorEnergySurfaces}). One may refer to this line as a Dirac exceptional line (node) following the terminology of a Dirac line node in condensed matter systems \cite{Young,Hong}.

In fact, a two-band model cannot host a Dirac EP in a two-dimensional parameter space, when the perturbation is just first-order. We note that such two-band models are widely used to generate Dirac cones in Hermitian systems \cite{Zhen}, and hence this finding highlights another difference between Hermitian and non-Hermitian systems in terms of their Dirac points. To show this difference, we note that all two-band models with an EP can be put into the Jordan normal form after a similar transformation. Therefore, we can take
\be
H_0=\begin{pmatrix}
0 & 1 \\
0 & 0
\end{pmatrix}
\ee
without loss of generality. We then express the perturbation as 
\be
\Delta H = \Delta_+ \sigma_+ + \Delta_-\sigma_- + \Delta_3 \sigma_3,\label{eq:2band_Pauli}
\ee
where we have neglected perturbations proportional to the identity matrix because they merely cause a shift of the whole spectrum. Here $\sigma_\pm = (\sigma_1 \pm i\sigma_2)/2$ and $\sigma_i\,(i=1,2,3)$ are the three Pauli matrices. $\Delta_\pm,\Delta_3\in\mathbb{C}$ are three complex perturbation amplitudes of the same order. The two energy eigenvalues are then given by
\be
\omega_\pm = \pm\sqrt{4\Delta_-(1+\Delta_+) + \Delta_3^2}.
\ee
Clearly, due to the leading order term $4\Delta_-$ in the randicand, the dispersion of this system cannot be conical; only when $\Delta_-$ is second order (i.e., $\Delta_-=\delta_-^2\sim\Delta_3^2$) do we recover a conic dispersion to the leading order:
\be
\omega_\pm \approx \pm\sqrt{4\delta_-^2 + \Delta_3^2}\quad(\delta_-,\Delta_3\in\mathbb{R}). \label{eq:DiracCone_2band}
\ee
In comparison, $H_0$ in a two-band Hermitian model with a Dirac point would simply vanish, and we have 
\be
\omega_\pm = \pm\sqrt{4\Delta_-\Delta_+ + \Delta_3^2} 
\ee
instead. A Dirac cone is then found, e.g., by letting $\Delta_3\in\mathbb{R}$ together with either $\Delta_-=\Delta_+\in\mathbb{R}$ or $\Delta_-=-\Delta_+\in i\mathbb{R}$.

\section{Imaginary Dirac cone}

If we multiply a non-Hermitian Hamiltonian with a Dirac EP by $i$, it is clear that the Dirac cone now exists in the imaginary part of the energy, which is uniquely non-Hermitian. Moreover, the general analysis in the last section, particular Eq.~(\ref{eq:DiracCone_2band}), indicates that it is unnecessary to change the unperturbed Hamiltonian $H_0$ to construct an imaginary Dirac cone with an EP at its center; we just need to change $\delta_-,\Delta_3$ from real to imaginary in this two-band model. 

\begin{figure}[tb]
\includegraphics[clip,width=\linewidth]{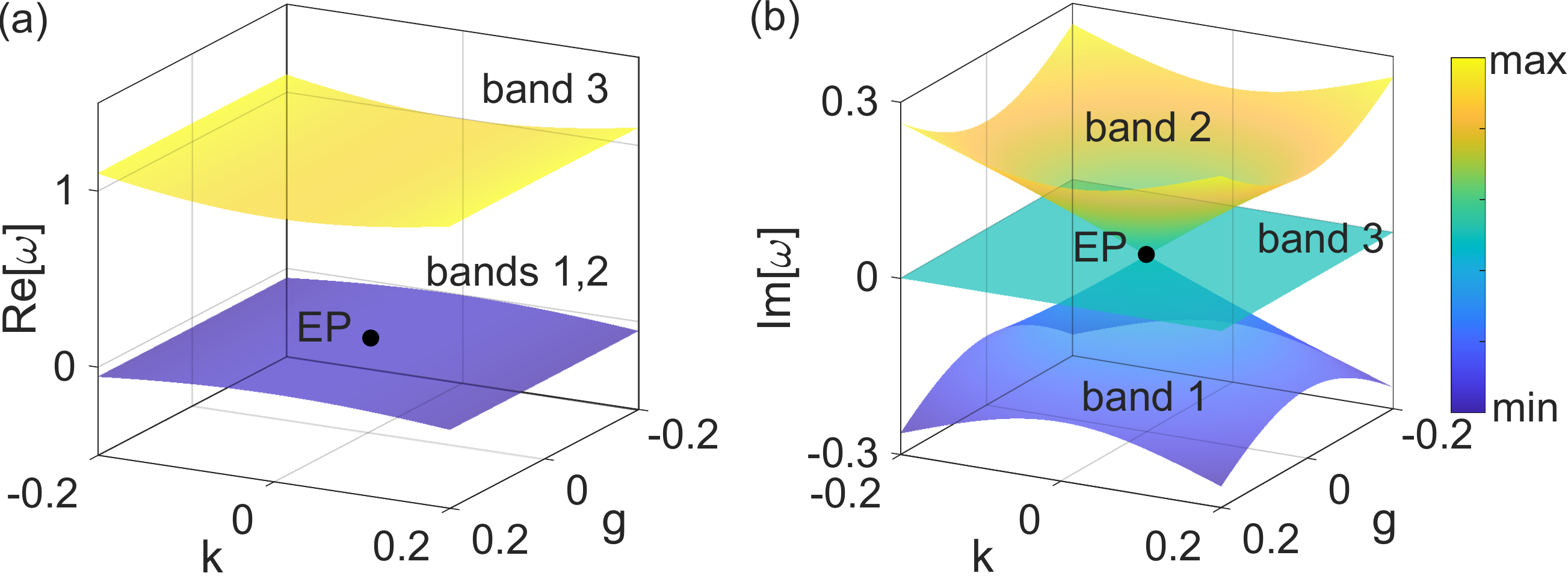}
\caption{Imaginary Dirac cone with an EP at its center. (a) and (b) show the real and imaginary parts of the three energies of the Hamiltonian given in Eq.~(\ref{eq:imagDiracCone}).  
} \label{fig:imagDiracEP}
\end{figure}

In a three-band model, we find that the following Hamiltonian hosts an imaginary Dirac cone:
\be
H = \begin{pmatrix}
ik & g & 1 \\
g  & 1 & g \\
0  & g & -ik   
\end{pmatrix}\quad (k,g\in\mathbb{R}).\label{eq:imagDiracCone}
\ee
The unperturbed Hamiltonian still has real eigenvalues $0$ (EP) and $1$. The real and imaginary parts of the three energy eigenvalues are shown in Fig.~\ref{fig:imagDiracEP}, where the real parts of the two coalesced eigenvalues at the EP stay the same in the two-dimensional parameter space. In other words, these two bands are complex conjugates, with the other band being real. These observations are consistent with the characteristic equation
\be
\omega^3-\omega^2+(k^2-2g^2)\omega-(k^2+g^2)=0,
\ee
which again has real coefficients. When expanding near the EP, we can drop the higher-order cubic term and solve the remaining quadratic equation. The result is 
\be
\omega_\pm = \pm i\sqrt{k^2+g^2}
\ee
to the leading order, showing the explicit conical and linear dispersion.

\section{Conclusion and discussions}

In summary, we studied a novel type of non-Hermitian degeneracies around which the energy spectrum remains real, while they manifest eigenvector coalescence in accordance with the definition of EPs. These \textit{Dirac EPs} are characterized by a conical dispersion around them, for which the standard perturbative description using the alternating Puiseux series fails. We also identified its imaginary counterpart, where a Dirac cone is formed in the imaginary parts of the energies with an EP at its center.

The Dirac EPs we presented emerge in a non-Hermitian three-band model, which can can be reduced to two-band models in several ways. This reduction may result in either Hermitian or non-Hermitian models which are \textit{isospectral}, and both host degeneracy points. This observation led to a startling discovery: There are Hermitian and non-Hermitian systems that have the same real-valued energy spectrum in their entire two-dimensional parameter space, and their degeneracy points consist of diabolic points in the former, and Dirac EPs in the latter. While isospectral non-Hermitian and Hermitian systems have been reported before (see, for example, Ref.~\cite{DiracEP0}), one with EPs and existing in an entire two-dimensional parameter space has not. As mentioned, we were able to achieve this isospectral property thanks to the elimination of conventional EPs, whose branch cuts would make the non-Hermitian spectrum complex in that region of the two-dimensional parameter space.

Finally, we showed that for two-band models, it is impossible to generate Dirac EPs with linear dispersion along all directions in a two-dimensional parameter space, if only first-order perturbations are introduced to an underlying non-Hermitian Hamiltonian at the EP; second-order terms are necessary to produce Dirac EPs with a conical dispersion. We note that this result holds for higher-dimensional systems as well, as our derivation based on Eq.~(\ref{eq:2band_Pauli}) is independent of the physical dimensions.
This property is in stark contrast to Hermitian systems, where the conical dispersion around a Dirac point is produced with first-order perturbations. %with higher-order perturbations leading to a higher Chern number once a gap is opened \cite{Fang}. 
This observation represents another intriguing difference between Dirac points in Hermitian and non-Hermitian systems. 

The results in this work broaden our understanding of non-Hermitian degeneracies. We break the traditional link between eigenvector coalescence and the characteristic integer root dispersion of the eigenvalue spectrum around the exceptional point in a two-dimensional parameter space. While a similar finding was known in the mathematical literature \cite{Ma}, it was based on a single (complex) parameter turning, where the real and imaginary parts of the perturbed Hamiltonian are collinear. When two independent perturbations are allowed instead as we do here (i.g., either using $\Delta\tau,k$ or $\alpha,k$), this result was found to break down in general and no traces of a Dirac EP were found \cite{EP3}: The dispersion near the EP is no longer linear except for one particular direction \cite{Fu}, consistent with previous findings \cite{Ma}. Furthermore, although the enhanced sensitivity stemming from the nonlinear dispersion of an EP is useful for applications in sensing, it simultaneously imposes a challenge for tuning and observing exceptional points. This fact, combined with the complex character of the non-Hermitian spectrum, often obscure the physics of eigenvector coalescence at the EP. Our work provides a more tolerant platform towards studying eigenstates coalescence in a two-dimensional parameter space, which may be generalized to higher dimensions as well.

This project is supported by NSF under Grant No. PHY-1847240 and and ECCS-1846766.

\end{document}